\documentclass[twocolumn,preprintnumbers,a4paper,
superscriptaddress,floatfix,twoside]{revtex4}

\usepackage{bm}
\usepackage{epsfig}
\usepackage{graphics}
\usepackage{natbib}

\usepackage{fancyh}
\usepackage[l]{floatflt}
\usepackage{epsfig}
\usepackage{amssymb}
\usepackage{latexsym}
\usepackage{times}
\usepackage{slashed}
\usepackage{upgreek}
\usepackage{amsmath}
\usepackage{amsfonts}
\usepackage{amsbsy}
\usepackage{amscd}
\usepackage{bbm}

\newcommand{\eec}{\end{center}}
\newcommand{\bec}{\begin{center}}

\newcommand{\eem}{\end{matrix}}
\newcommand{\bem}{\begin{matrix}}
\newcommand{\eeq}{\end{equation}}
\newcommand{\beq}{\begin{equation}}
\newcommand{\ba}{\begin{array}}
\newcommand{\ea}{\end{array}}
\newcommand{\bea}{\begin{eqnarray}}
\newcommand{\eea}{\end{eqnarray}}
\newcommand{\baq}{\begin{eqnarray}}
\newcommand{\eaq}{\end{eqnarray}}
\newcommand{\beqs}{\begin{subequations}}
\newcommand{\eeqs}{\end{subequations}}

\newcommand\eqs[2]{Eqs.~(\ref{#1}) and (\ref{#2})}

\newcommand{\ftn}{\footnotesize}

\newcommand{\TeV}{{\mbox{\rm TeV}}}

\newcommand{\GeV}{{\mbox{\rm GeV}}}

\newcommand{\sFref}[2]{Fig.~\ref{#1}-{\small \sf ({#2})}}

\newcommand{\etal}{{\it et al.\/}}

\def\to{\rightarrow}

\def\lf{\left(}
\def\rg{\right)}
\newcommand\vev[1]{\langle {#1} \rangle}

\newcommand{\Nhi}{\ensuremath{N_{\rm HI*}}}
\newcommand{\Gsm}{\ensuremath{G_{\rm SM}}}
\newcommand{\Nr}{\ensuremath{{\sf N}}}
\newcommand{\Vhi}{\ensuremath{V_{\rm HI}}}

\newcommand{\Whi}{\ensuremath{W_{\rm HI}}}

\newcommand{\Vhio}{\ensuremath{V_{\rm HI0}}}

\newcommand{\mP}{\ensuremath{m_{\rm P}}}

\newcommand{\Mgut}{\ensuremath{M_{\rm GUT}}}
\newcommand{\Ggut}{\ensuremath{G_{\rm GUT}}}
\newcommand{\Gfl}{\ensuremath{G_{\rm 5_X}}}
\newcommand{\Glr}{\ensuremath{G_{\rm LR}}}

\newcommand{\ns}{\ensuremath{n_{\rm s}}}
\newcommand{\as}{\ensuremath{\alpha_{\rm s}}}

\newcommand{\Dex}{\ensuremath{\Delta_{\rm m*}}}

\newcommand{\Trh}{\ensuremath{T_{\rm rh}}}
\newcommand{\sg}{\ensuremath{\sigma}}
\newcommand{\sgex}{\ensuremath{\sigma_*}}

\newcommand{\kp}{\ensuremath{\kappa}}
\newcommand{\ks}{\ensuremath{k_{4S}}}
\newcommand{\kss}{\ensuremath{k_{6S}}}
\newcommand{\ksss}{\ensuremath{k_{8S}}}

\newcommand{\kst}{\ensuremath{k_{10S}}}
\newcommand{\ksv}{\ensuremath{k_{12S}}}

\newcommand{\ck}{\ensuremath{c_{2K}}}
\newcommand{\ckk}{\ensuremath{c_{4K}}}
\newcommand{\ckx}{\ensuremath{c_{6K}}}
\newcommand{\ckh}{\ensuremath{c_{8K}}}

\newcommand{\hepph}[1]{{\ftn\tt hep-ph/#1}}

\newcommand{\astroph}[1]{{\ftn\tt astro-ph/#1}}
\newcommand{\arxiv}[1]{{\ftn\tt  arXiv:#1}}

\newcommand{\Eref}[1]{Eq.~(\ref{#1})}
\newcommand{\Sref}[1]{Sec.~\ref{#1}}
\newcommand{\Fref}[1]{Fig.~\ref{#1}}
\newcommand{\Tref}[1]{Table~\ref{#1}}
\newcommand{\cref}[1]{Ref.~\cite{#1}}

\def\Ka{K\"{a}hler potential~}
\def\Kap{K\"{a}hler potential}
\def\p{|S|}

\newcommand{\bdhh}{{\ensuremath{\normalsize I{\kern-2.9pt H}}}}

\renewcommand{\refname}{{\bf\scshape References}}

\renewenvironment{subequations}{%
\refstepcounter{equation}%
\setcounter{parentequation}{\value{equation}}%
  \setcounter{equation}{0}
  \ignorespaces
}{%
  \setcounter{equation}{\value{parentequation}}%
  \ignorespacesafterend
}

\begin{document}


\title{\boldmath\bf\scshape  Upper Bound on the Tensor-to-Scalar Ratio
\\ in GUT-Scale Supersymmetric Hybrid Inflation}

\author{\scshape  Matthew Civiletti}
\affiliation{ Bartol Research Institute, Department of Physics and
Astronomy, University of Delaware, Newark, DE 19716, USA\\  {\sl
e-mail addresses: }{\ftn\tt mcivil@bartol.udel.edu,
shafi@bartol.udel.edu}}
\author{\scshape Constantinos Pallis}
\affiliation{Departament de F\'isica Te\`orica and IFIC,
Universitat de Val\`encia-CSIC, E-46100 Burjassot, SPAIN
\\  {\sl e-mail address: }{\ftn\tt cpallis@ific.uv.es}}
\author{\scshape  Qaisar Shafi} 
\affiliation{ Bartol Research Institute, Department of Physics and
Astronomy, University of Delaware, Newark, DE 19716, USA\\  {\sl
e-mail addresses: }{\ftn\tt mcivil@bartol.udel.edu,
shafi@bartol.udel.edu}}


\begin{abstract}

\noindent {\ftn \bf\scshape Abstract:} We explore the upper bound
on the tensor-to-scalar ratio $r$ in supersymmetric (F-term)
hybrid inflation models with the gauge symmetry breaking scale set
equal to the value $2.86\cdot10^{16}~{\rm GeV}$, as dictated by
the unification of the MSSM gauge couplings. We employ a unique
renormalizable superpotential and a quasi-canonical K\"ahler
potential, and the scalar spectral index $n_s$ is required to lie
within the two-sigma interval from the central value found by the
Planck satellite. In a sizable region of the parameter space the
potential along the inflationary trajectory is a monotonically
increasing function of the inflaton, and for this case,
$r\lesssim2.9\cdot10^{-4}$, while the spectral index running,
$|d\ns/d\ln k|$, can be as large as $0.01$. Ignoring higher order
terms which ensure the boundedness of the potential for large
values of the inflaton, the upper bound on $r$ is significantly
larger, of order $0.01$, for subplanckian values of the inflaton,
and $|d\ns/d\ln k|\simeq0.006$.
\\ \\ {\scriptsize {\sf PACs numbers: 98.80.Cq, 12.60.Jv}


}

\end{abstract}\pagestyle{fancyplain}

\maketitle

\rhead[\fancyplain{}{ \bf \thepage}]{\fancyplain{}{\sl Upper Bound
on the Tensor-to-Scalar Ratio in GUT-Scale SUSY Hybrid Inflation}}
\lhead[\fancyplain{}{\sl \leftmark}]{\fancyplain{}{\bf \thepage}}
\cfoot{}

\section{Introduction}

\emph{Supersymmetric} ({\small\sf SUSY}) hybrid inflation based on
F-terms, also referred to as \emph{F-term hybrid inflation} ({\sf
\small FHI}), is one of the simplest and well-motivated
inflationary models \cite{susyhybrid,hybrid}. It is tied to a
renormalizable superpotential uniquely determined by a global
$U(1)$ R-symmetry, does not require fine tuned parameters and it
can be naturally followed by the breaking of a \emph{Grand Unified
Theory} ({\sf \small GUT}) gauge symmetry, such as $G_{B-L}=
G_{\rm SM}\times U(1)_{B-L}$ \cite{bl}, where ${G_{\rm SM}}=
SU(3)_{\rm C}\times SU(2)_{\rm L}\times U(1)_{Y}$ is the gauge
group of the \emph{Standard Model} ({\small\sf SM}),
$\Glr=SU(3)_{\rm C}\times SU(2)_{\rm L} \times SU(2)_{\rm R}
\times U(1)_{B-L}$ \cite{dvali, vlachos}, and flipped $SU(5)$
\cite{barr, ant, davis, flipped}, with gauge symmetry
$\Gfl=SU(5)\times U(1)_X$. Let us clarify, in passing, that the
term ``GUT'' is used in the sense of the gauge coupling
unification within \emph{Minimal SUSY SM} ({\sf \small MSSM}),
although the aforementioned gauge groups are not simple. Such
models can arise from string compactifications, see for e.g.
\cref{tracas,ant}. The embedding of the simplest model of FHI
within a higher gauge group may suffer from the production of
cosmic defects which can be evaded, though, in the cases of smooth
\cite{smooth} or shifted \cite{shifted} FHI.

In the simplest implementation of FHI \cite{susyhybrid}, we should
note that the potential along the inflationary track is completely
flat at tree level. The inclusion of \emph{radiative corrections}
({\sf \small  RCs}) \cite{susyhybrid} produce a slope which is
needed to drive inflaton towards the SUSY vacuum. In this
approximation the predicted scalar spectral index $n_s\simeq0.98$,
is in slight conflict with the latest WMAP \cite{wmap} and PLANCK
\cite{plin} data based on the standard power-law cosmological
model with \emph{Cold Dark Matter and a cosmological constant}
({\sf \small $\Lambda$CDM}). Furthermore, the gauge symmetry
breaking scale $M$ turns out to be close to (but certainly lower
than) its SUSY value, $\Mgut\simeq2.86\cdot10^{16}~{\rm GeV}$.

A more complete treatment which incorporates \emph{supergravity}
({\sf \small SUGRA}) corrections \cite{senoguz} with canonical
(minimal) \Kap, as well as an important soft SUSY breaking term
\cite{sstad,mfhi}, has been shown to yield values for $\ns$ that
are fully compatible with the data \cite{wmap,plin}, with $M$ in
this case somewhat lower than the one obtained in
\cref{susyhybrid}. A reduction of $M$ is certainly welcome if FHI
is followed by the breaking of an abelian gauge symmetry, since it
helps to reconcile $M$ with the bound \cite{plcs} placed on it by
the non-observation of cosmic strings \cite{jp, gmb, mfhi, buch}.

The minimal FHI scenario described above, while perfectly
consistent with the current observations, requires some
modification if one desires to incorporate values of $M$ that are
comparable or equal to $\Mgut$. This is indispensable in cases
where $\Ggut$ includes non abelian factors besides $\Gsm$, which
are expected to disturb the successful gauge coupling unification
within MSSM. In this letter, we would like to emphasize that the
observationally favored values (close to 0.96) for $\ns$ with $M$
equal to the SUSY GUT scale can be readily achieved within FHI by
invoking a specific type of non-minimal \Kap, first proposed in
\cref{rlarge}. In particular, a convenient choice of the
next-to-minimal and the next-to-next-to-minimal term of the
adopted \Ka generates \cite{rlarge,hinova,alp} a positive mass
(quadratic) term for the inflaton and a sizeable negative quartic
term which assist us to establish FHI of hilltop type \cite{lofti}
in most of the allowed parameter space of the model. Our
objectives can also be achieved in smaller regions of the allowed
parameter space even with monotonic inflationary potential and
therefore complications related to the initial conditions of FHI
can be safely eluded. Acceptable $\ns$ values within this set-up
are accompanied with an enhancement of the running of $\ns$,
$\as$, and the scalar-to-tensor ratio, $r$, which reach, thereby,
their maximal possible values within FHI if we take into account
that $M$'s larger than $\Mgut$ are certainly less plausible. Note,
in passing, that the reduction of $\ns$ by generating a negative
mass (quadratic) term for the inflaton, as done in \cref{gpp}, is
not suitable for our purposes since $M$ remains well below
$\Mgut$.

Below, we briefly review in \Sref{fhi} the basics of FHI when it
is embedded in nonminimal SUGRA and recall in \Sref{fhi3} the
observational and theoretical constraints imposed on our model. In
\Sref{res} we exhibit our updated results, and our conclusions are
summarized in \Sref{con}.

\section{FHI With Nonminimal K\"{a}hler Potential}\label{fhi}

\paragraph{\sf\scshape\small Spontaneous Breaking of $\Ggut$.}\label{fhi1}
The standard FHI can be realized by adopting the superpotential
\beq W = \kappa S\left(\bar \Phi\Phi-M^2\right)\label{Whi}\eeq
which is the most general renormalizable superpotential consistent
with a continuous R-symmetry \cite{susyhybrid} under which \beq S\
\to\ e^{i\alpha}\,S,~\bar\Phi\Phi\ \to\ \bar\Phi\Phi,~W \to\
e^{i\alpha}\, W.\label{Rsym} \eeq Here $S$ is a $\Ggut$-singlet
left-handed superfield, and the parameters $\kappa$ and $M$ are
made positive by field redefinitions. In our approach
$\bar{\Phi}$, $\Phi$ are identified with a pair of left-handed
superfields conjugate under $\Ggut$ which break  $\Ggut$ down to
$\Gsm$. Indeed,  along the D-flat direction $|\bar\Phi|=|\Phi|$
and the SUSY potential, $V_{\rm SUSY}$, extracted (see e.g.
ref.~\cite{lectures, hinova}) from $W$ in Eq.~(\ref{Whi}), reads
\beq \label{VF} V_{\rm SUSY}=
\kappa^2\left((|\Phi|^2-M^2)^2+2|S|^2 |\Phi|^2\right).\eeq
From $V_{\rm SUSY}$ in Eq.~(\ref{VF}) we find that the SUSY vacuum
lies at
\beq
\vev{S}=0\>\>\>\mbox{and}\>\>\>\left|\vev{\Phi}\right|=\left|\vev{\bar\Phi}\right|=M,
\label{vevs} \eeq
where the vacuum expectation values of $\Phi$ and $\bar\Phi$ are
developed along their SM singlet type components. As a
consequence, $\Whi$ leads to the spontaneous breaking of $\Ggut$
to $\Gsm$. We single out the following two cases:

\begin{itemize}

\item $\Ggut=\Glr$ where $\Phi$ and $\bar\Phi$ belong to the
$({\bf 1, 1, 2}, -1)$ and $({\bf 1, 1, \bar 2}, 1)$ representation
of $\Glr$ -- cf. \cref{vlachos,alp}. The symmetry breaking in this
case is
$$ SU(2)_{\rm R}\times U(1)_{B-L} \to U(1)_{Y}.$$ Therefore, 3 of
the 4 generators of $SU(2)_{\rm R}\times U(1)_{B-L}$ are broken,
leading to 3 Goldstone bosons which are absorbed by the 3 gauge
bosons which become massive. Among them, $W_{\rm R}^\pm$ with
masses $m_{W_{\rm R}^\pm} = gM$ correspond to the charged
$SU(2)_R$ gauge generators, and one, $A$, to a linear combination
of the $SU(2)_R$ and $U(1)_{B-L}$ generator with mass $m_A =
\sqrt{5/2}gM$, where $g$ is the SUSY  gauge coupling constant at
the GUT scale.

\item $\Ggut=\Gfl$, where $\Phi$ and $\bar\Phi$ belong to the
$({\bf 10}, 1)$ and $({\bf \overline{10}}, -1)$ representation of
$\Gfl$ -- cf. \cref{ant,davis,flipped}. In this case, 13 of the 25
generators of $\Gfl$ are broken, giving rise via the Higgs
mechanism to 13 massive gauge bosons. In particular, 12 gauge
bosons which correspond to the generators of $SU(5)$ acquire
masses $m_{X_i^\pm}=m_{Y_i^\pm} = gM$, and one gauge boson
associated with a linear combination of the $SU(5)$ and $U(1)_{X}$
generators acquires a mass $m_A = \sqrt{32/34}gM$ -- cf.
\cref{davis}.

\end{itemize}

In both cases no topological defects are generated during the
breaking of $\Ggut$, in contrast to gauge groups such as
$SU(4)_{\rm C}\times SU(2)_{\rm L}\times SU(2)_{\rm R}$, $SU(5)$
or $SO(10)$ which lead to the production of magnetic monopoles.

\paragraph{\sf\scshape \small The Inflationary Stage.}
The superpotential $\Whi$ in Eq.~(\ref{Whi}) gives rise to FHI
since, for large enough values of $|S|$, there exist a flat
direction
\begin{equation} \label{V0}\bar\Phi={\Phi}=0 ~~\mbox{where}~~V_{\rm SUSY}\lf{\Phi}=0\rg=V_{\rm HI0}=\kappa^2 M^4.\eeq
Obviously, $V_{\rm HI0}$ provides us with a constant potential
energy density which can be used to drive inflation. The
realization of FHI in the context of SUGRA requires a specific
\Kap. We consider here a fairly generic form of the K\"{a}hler
potential, which does not deviate much from the canonical one
\cite{senoguz,mfhi}; further it respects the R symmetry of
\Eref{Rsym}. Namely we take
\bea\nonumber K&=&\p^2+|\Phi|^2+|\bar
\Phi|^2+{1\over4}\ks{\p^4\over\mP^2}+{1\over6}\kss{\p^6\over\mP^4}
\\&+&{1\over8}\ksss{\p^8\over\mP^6}
+{1\over10}\kst{\p^{10}\over\mP^8}
+{1\over12}\ksv{\p^{12}\over\mP^{10}}+\cdots~~~~\label{K} \eea
where $\ks,\kss,\ksss,\kst$ and $\ksv$ are positive or negative
constants of order unity and the ellipsis represents higher order
terms involving the waterfall fields ($\bar \Phi$ and $\Phi$) and
$S$. We can neglect these terms since they are irrelevant along
the inflationary path. Finally, we include the RCs. These
originate from a mass splitting in the $\Phi-\bar{\Phi}$
supermultiplets, caused by SUSY breaking along the inflationary
valley \cite{susyhybrid}. We end up with the following
inflationary potential -- see e.g. \cref{hinova,alp}:
\beq\label{Vol} V_{\rm HI}\simeq V_{\rm HI0}\left(1+c_{\rm
HI}+\sum_{\nu=1}^5(-1)^{\nu}c_{2\nu
K}\lf{\sg\over\sqrt{2}\mP}\rg^{2\nu}\right),\eeq
where $\sigma=\sqrt{2}\p$ is the canonically (up to the order
$\p^2$) normalized inflaton field. The contribution of RCs read
\beqs\beq \label{Vcor}c_{\rm HI}=
{\kappa^2 \Nr\over 32\pi^2}\left(2 \ln {\kappa^2x M^2 \over Q^2}
+f_{\rm rc}(x)\right),\eeq
where $\Nr$, for our cases, is the dimensionality of the
representations to which $\bar{\Phi}$ and $\Phi$ belong. We have
$\Nr=2$ [$\Nr=10$] when $\Ggut=\Glr$ [$\Ggut=\Gfl$].  Also $Q$ is
a renormalization scale, $x=\sigma^2/2M^2$, and
\beq f_{\rm rc}(x)=(x+1)^{2}\ln\lf1+{1/
x}\rg+(x-1)^{2}\ln\lf1-{1/x}\rg.\label{frc}\eeq\eeqs The remaining
coefficients, $c_{2\nu K}$, in \Eref{Vol} can be expressed as
functions of the $k$'s in \Eref{K} \cite{hinova,alp}. From them
only the first two play a crucial role during the inflationary
dynamics; they are
%
%
\beq\label{cks} c_{2K}=\ks~~\mbox{and}~~ c_{4K}={1\over2} - {7
\ks\over4} + \ks^2 - {3\kss\over2}\eeq
The residual higher order terms in the expansion of \Eref{Vol}
prevent a possible runaway behavior of the resulting $\Vhi$ -- see
point 8 of \Sref{fhi3}. For completeness, we include also them:
\beqs\bea c_{6K}&=&-\frac{2}{3} + \frac{3 \ks}{2} - \frac{7
\ks^2}{4} + \ks^3 + \frac{10\kss}{3}\nonumber \\&& - 3 \ks\kss +
2\ksss,\label{c6k}\eea\bea c_{8K}&=& \frac{3}{8} - \frac{5
\kst}{2} - \frac{13 \ks}{24} + \frac{41 \ks^2}{32} - \frac{7
\ks^3}{4} + \ks^4 \nonumber
\\&&- \frac{13\kss}{4} + \frac{143 \ks\kss}{24} - \frac{9 \ks^2\kss}{2} +
\frac{9\kss^2}{4}\nonumber \\&& - \frac{39\ksss}{8} + 4 \ks \ksss,
\label{c8k}\\ c_{10K}&=&-\frac{2}{15} + \frac{32 \kst}{5} + 3 \ksv
+ \frac{\ks}{24} - 5 \kst \ks  \nonumber\\&& - \frac{13 \ks^2}{24}
+ \frac{41 \ks^3}{32} - \frac{7 \ks^4}{4} + \ks^5 +
\frac{5\kss}{3} \nonumber\\&& - \frac{29 \ks\kss}{6} + \frac{103
\ks^2\kss}{12} - 6 \ks^3\kss - 5\kss^2 \nonumber \\ && + \frac{27
\ks \kss^2}{4} + 5\ksss - \frac{67 \ks\ksss}{8} \nonumber \\&& + 6
\ks^2\ksss- 6\kss \ksss. \label{c10k}\eea\eeqs

Let us note, lastly, that the most important contribution
\cite{sstad} to $\Vhi$ from the soft SUSY breaking terms of the
order of $(1-10)~\TeV$ does not play any essential role in our
set-up due to large $M$'s employed here -- cf.~\cref{mfhi}.

\section{Constraining the Model Parameters}\label{fhi3}

Under the assumptions that (i) the observed curvature perturbation
is generated wholly by $\sigma$ and (ii) FHI is followed in turn
by matter and radiation era, our inflationary set-up can be
qualified by imposing a number of observational (1-3) and
theoretical (4-8) requirements specified below:

\setcounter{paragraph}{0}

\paragraph{} The number of e-foldings that the scale $k_*=0.05/{\rm Mpc}$
undergoes during FHI is at least enough to resolve the horizon and
flatness problems of standard Big Bang cosmology. Employing
standard methods \cite{hinova, plin}, we can derive the relevant
condition:
\begin{equation}  \label{Nhi}
N_{\rm HI*}=\int_{\sigma_{\rm f}}^{\sigma_{*}}\,
\frac{d\sigma}{m^2_{\rm P}}\: \frac{V_{\rm HI}}{V'_{\rm
HI}}\simeq19.4+{2\over 3}\ln{V^{1/4}_{\rm HI0}\over{1~{\rm GeV}}}+
{1\over3}\ln {T_{\rm rh}\over{1~{\rm GeV}}},
\end{equation}
where the prime denotes derivation w.r.t. $\sigma$, $\sigma_{*}$
is the value of $\sigma$ when $k_*$ crosses outside the horizon of
FHI, and $\sigma_{\rm f}$ is the value of $\sigma$ at the end of
FHI. This coincides with either the critical point $\sigma_{\rm
c}=\sqrt{2}M$ appearing in the particle spectrum of the
$\Phi-\bar\Phi$ system during FHI -- see \Eref{frc} --, or the
value for which one of the slow-roll parameters \cite{review}
\beq \label{slow} \epsilon\simeq{m^2_{\rm P}}\left({V'_{\rm
HI}}/\sqrt{2}{V_{\rm HI}}\right)^2~~\mbox{and}~~\eta\simeq
m^2_{\rm P}~{V''_{\rm HI}}/{V_{\rm HI}} \eeq
exceeds unity. Since the resulting $\kp$ values are sizably larger
than $(M/\mP)^2$ -- see next section -- we do not expect the
production of extra e-foldings during the waterfall regime, which
in our case turns out to be nearly instantaneous -- cf.
\cref{bjorn}.

\paragraph{} The amplitude $A_{\rm s}$ of the power spectrum of the curvature
perturbation, which is generated during FHI and can be calculated
at $k_{*}$ as a function of $\sg_*$, must be consistent with the
data \cite{wmap, plin}, i.e.
\begin{equation} \label{Prob}
\sqrt{A_{\rm s}}= \frac{1}{2\sqrt{3}\, \pi m^3_{\rm P}}\;
\left.\frac{V_{\rm HI}^{3/2}(\sigma_*)}{|V_{{\rm
HI},\sigma}(\sigma_*) |}\right.\simeq\: 4.686\cdot 10^{-5}.
\end{equation}

\paragraph{} The (scalar) spectral index $n_{\rm s}$, its running,
$a_{\rm s}=d\ns/d\ln k$, and the scalar-to-tensor ratio, $r$,
given by
\beqs\bea \label{nS} && n_{\rm s}=1-6\epsilon_*\ +\ 2\eta_*,~~~~~~\\
&& \label{aS} \alpha_{\rm s}={2}\left(4\eta_*^2-(n_{\rm
s}-1)^2\right)/3-2\xi_*~~\mbox{and}~~ r=16\epsilon_*, ~~~~~~~
\eea\eeqs
where $\xi\simeq m_{\rm P}^4~V'_{\rm HI} V'''_{\rm HI}/V^2_{\rm
HI}$ and all the variables with the subscript $*$ are evaluated at
$\sigma=\sigma_{*}$, must be in agreement with the observational
data \cite{wmap, plin} derived in the framework of the
$\Lambda$CDM model:
\beqs\bea\label{nswmap} &&
\ns=0.9603\pm0.014~\Rightarrow~0.945\lesssim n_{\rm s}
\lesssim 0.975,~~~~~~\\
&&\label{obs3}\as=-0.0134\pm0.018~~\mbox{and}~~
r<0.11, \label{obs4}\eea\eeqs
at 95$\%$ \emph{confidence level} (c.l.). %
Limiting ourselves to $\as$'s consistent with the assumptions of
the power-law $\Lambda$CDM model, we further impose the following
upper bound: \beq |\as|\ll0.01,\label{aswmap} \eeq since, within
the cosmological models with running $\as$, $|\as|$'s of order
0.01 are encountered \cite{plin,wmap}.

\paragraph{} The $\Ggut$ breaking scale in \Eref{vevs} has to be
determined by the unification of the MSSM gauge coupling
constants, i.e.,
\beq \label{Mgut} gM\simeq2 \cdot 10^{16}~\GeV,\eeq
with $g\simeq0.7$ being the value of the unified gauge coupling
constant. Here $gM$ is the mass at the SUSY vacuum, \Eref{vevs},
of the non singlet under $\Gsm$ gauge bosons $W_{\rm R}^{\pm}$ if
$\Ggut=\Glr$ or $X^{\pm}$ and $Y^{\pm}$ if $\Ggut=\Gfl$ -- see
\Sref{fhi}.

\paragraph{} The expression
of $\Vhi$ in \Eref{Vol} is expected to converge at least for
$\sg\sim\sg_*$. This fact can be ensured if, for $\sg\sim\sg_*$,
each successive term $c_{2\nu K}$ in the expansion of $\Vhi$ (and
$K$) \Eref{Vol} (and \Eref{K}) is smaller than the previous one.
In practice, this objective can be easily accomplished if the
$k$'s in \Eref{K} -- or \Eref{Vol} -- are sufficiently low.

\paragraph{} It is reasonable to ask $\Vhi$ to be bounded from below as
$\sg\to\infty$. Given our ignorance, however, for the
pre-inflationary (i.e. for $\sg>\sg_*$) cosmological evolution we
do not impose this requirement as an absolute constraint.

\paragraph{} Depending on the values of the coefficients in \Eref{Vol},
$V_{\rm HI}$ is a either monotonic function of $\sigma$ or
develops a local minimum and maximum. The latter case may
jeopardize the implementation of FHI if $\sg$ gets trapped near
the minimum of $\Vhi$. It is, therefore, crucial to indicate the
regions where $V_{\rm HI}$ is a monotonically increasing function
of $\sigma$.

\paragraph{} Hilltop FHI
proceeds such that $\sigma$ rolls from $\sigma_{\rm max}$, which
is the point where the maximum of $V_{\rm HI}$ lies, down to
smaller values. Therefore a mild tuning of the initial conditions
is required \cite{gpp} in order to obtain acceptable $n_{\rm s}$
values, since for lower $n_{\rm s}$ values we must set $\sigma_*$
closer to $\sigma_{\rm max}$. We quantify the amount of tuning in
the initial conditions via the quantity \cite{gpp}:
\beq \Dex=\left(\sigma_{\rm max} - \sigma_*\right)/\sigma_{\rm
max}.\label{dms}\eeq
Large $\Dex$ values correspond to a more natural FHI scenario.

\section{Results}\label{res}

Our inflationary model depends on the parameters:
$$ \kp,~\ks,~\kss,~\ksss,~\kst,~\ksv,~\Nr,~\Trh,~~\mbox{and}~~\sigma_*\ ,$$
with $M$ fixed from \Eref{Mgut}. In our computation, we use as
input parameters $\ksss, ~\kst$ and $\ksv$. We also fix $T_{\rm
rh}\simeq10^{9}~\GeV$, which saturates the conservative gravitino
constraint and results in $\Nhi\simeq50$. Variation of $\Trh$ over
$1-2$ orders of magnitude is not expected to significantly alter
our findings -- see \Eref{Nhi}. We restrict $\kp$ and $\sigma_*$
such that Eqs.~(\ref{Nhi}) and (\ref{Prob}) are fulfilled. The
restrictions on $\ns$ from \Eref{nswmap} can be met by adjusting
$\ks$ and $\kss$, whereas the last three parameters of $K$ control
mainly the boundedness and the monotonicity of $\Vhi$; we thus
take them into account only if we impose restriction 6 of
\Sref{fhi3}. In these cases we set $\kst=-1$ and $\ksv=0.5$
throughout and we verify that these values do not play a crucial
role in the inflationary dynamics. We briefly comment on the
impact of the variation of $\ksss$ and $\Nr$ on our results. Using
Eq.~(\ref{aS}) we can extract $\alpha_{\rm s}$ and $r$.

Following the strategy of \cref{rlarge} we choose the sign of
$\ck=\ks$ to be negative -- cf. \cref{gpp}. As a consequence,
fulfilling of \Eref{nswmap} requires a negative $\ckk$ or positive
$\kss$ -- see \Eref{cks}. More explicitly, $\Vhi$ given by
Eq.~(\ref{Vol}) can be approximated as
\bea\label{Vnnm} \nonumber \Vhi&\simeq&\Vhio\,\left(1+c_{\rm HI}+|\ck|\frac{\sigma^2}{2\mP^2}-\,|\ckk|\frac{\sigma^4}{4\mP^4} \right.\\
&&\left.-\,|\ckx|\frac{\sigma^6}{8\mP^6}+\,|c_{8K}|\frac{\sigma^8}{16\mP^8}\right),\eea
and it may develop a non-monotonic behavior in a sizable portion
of the allowed parameter space. Employing \Eref{Vnnm}, we can show
that $\Vhi$ reaches a local maximum at the inflaton-field value:
\beqs\beq \label{sigmamax}\sigma_{\rm max}\simeq
\frac{\mP \sqrt{\pi|\ck| + \sqrt{\pi^2\ck^2 + \Nr\kp^2
|\ckk|}}}{\sqrt{2 \pi|\ckk|}},\end{equation} and a local minimum
at the inflaton-field value:
\begin{equation}
\label{sigmamin} \sigma_{\rm min}\simeq \mP\frac{\sqrt{3 |\ckx| +
\sqrt{9 \ckx^2 + 32 |\ckk \ckx|}}}{2 \sqrt{|\ckh|}}\cdot\eeq\eeqs
In deriving \Eref{sigmamax} we keep terms up to the fourth power
of $\sg$ in \Eref{Vnnm}, whereas for \Eref{sigmamin} we focus on
the last three terms of the expansion in the right-hand side of
\Eref{Vnnm}. For this reason, the latter result is independent of
$c_{\rm HI}$ and $\ck$.

\begin{figure}[!t]
\includegraphics[width=60mm,angle=-90]{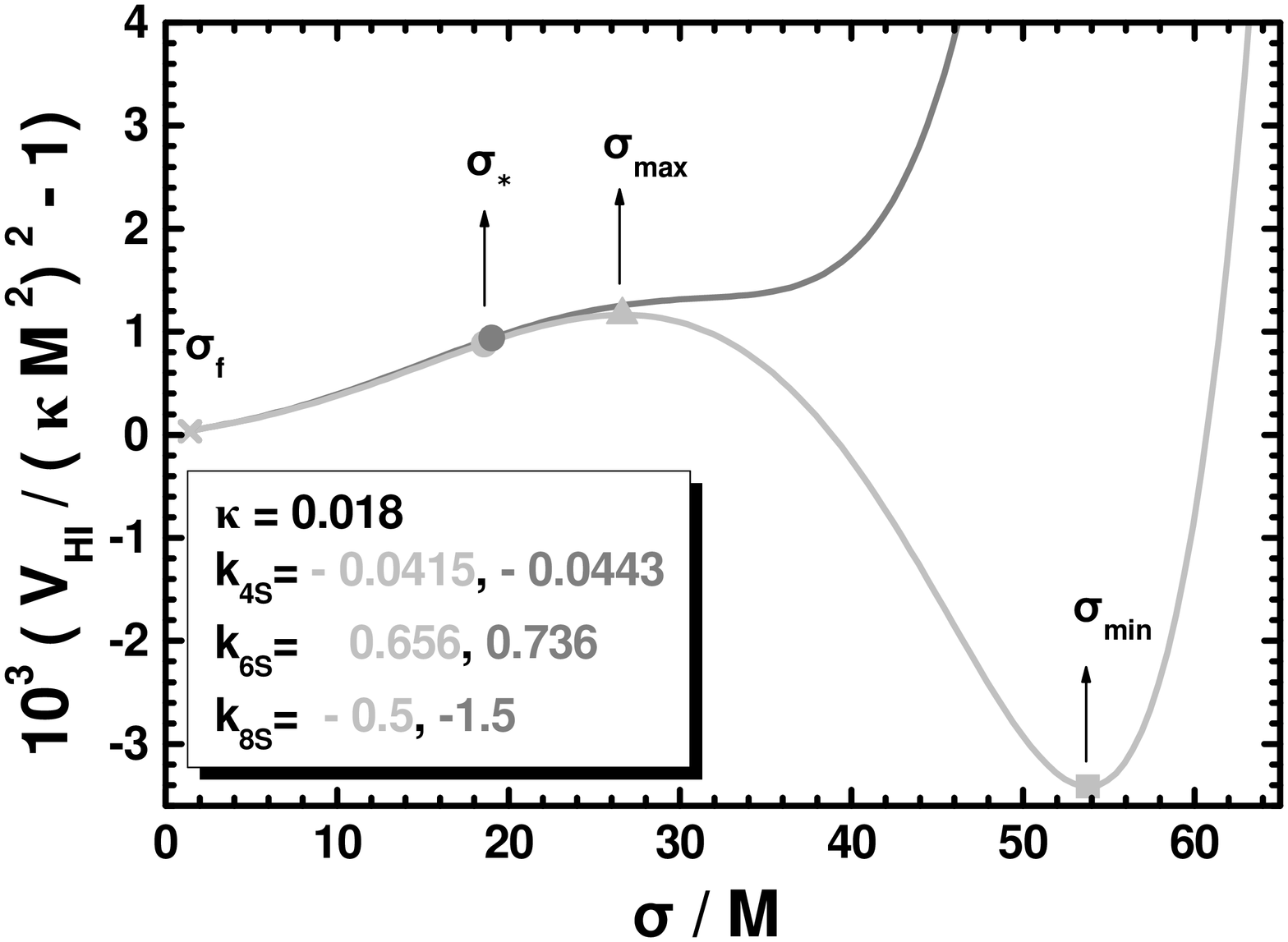}
\vspace*{-0.5cm} \caption{\sl The variation of $\Vhi$ in
\Eref{Vnnm} as a function of $\sg$ for $\ns=0.96$ taking $\Nr=10$,
$\kappa=0.018$, $\kst=-1,\ksv=0.5$ and
$\ks=-0.0443,\kss=0.736,\ksss=-1.5$
[$\ks=-0.0415,\kss=0.656,\ksss=-0.5$ ] (gray [light gray] line).
The values of $\sigma_*, \sigma_{\rm f}, \sg_{\rm max}$ and
$\sg_{\rm min}$ are also depicted.}\label{Vhi}
\end{figure}

\begin{figure*}[!t]
\centering
\includegraphics[width=60mm,angle=-90]{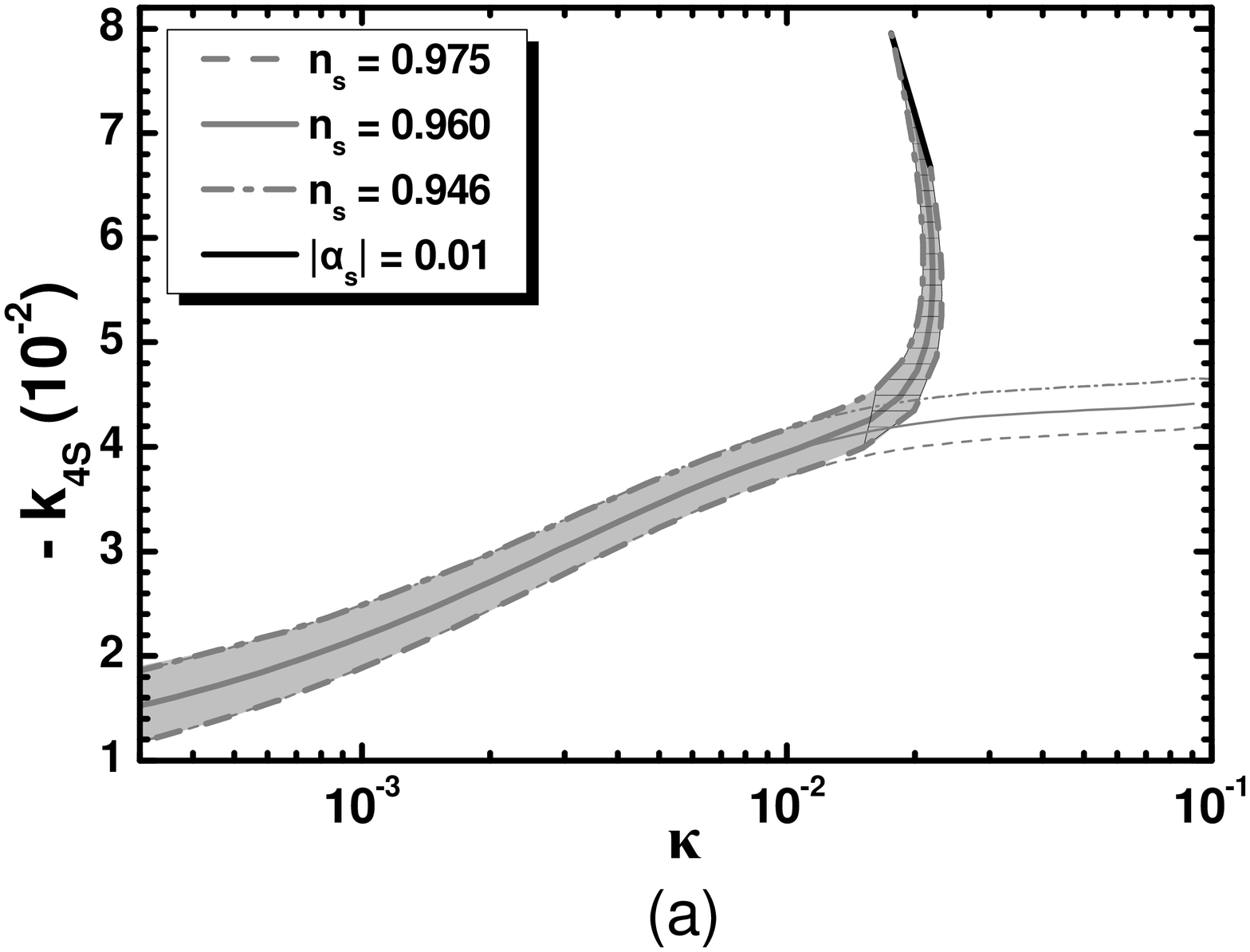}
\includegraphics[width=60mm,angle=-90]{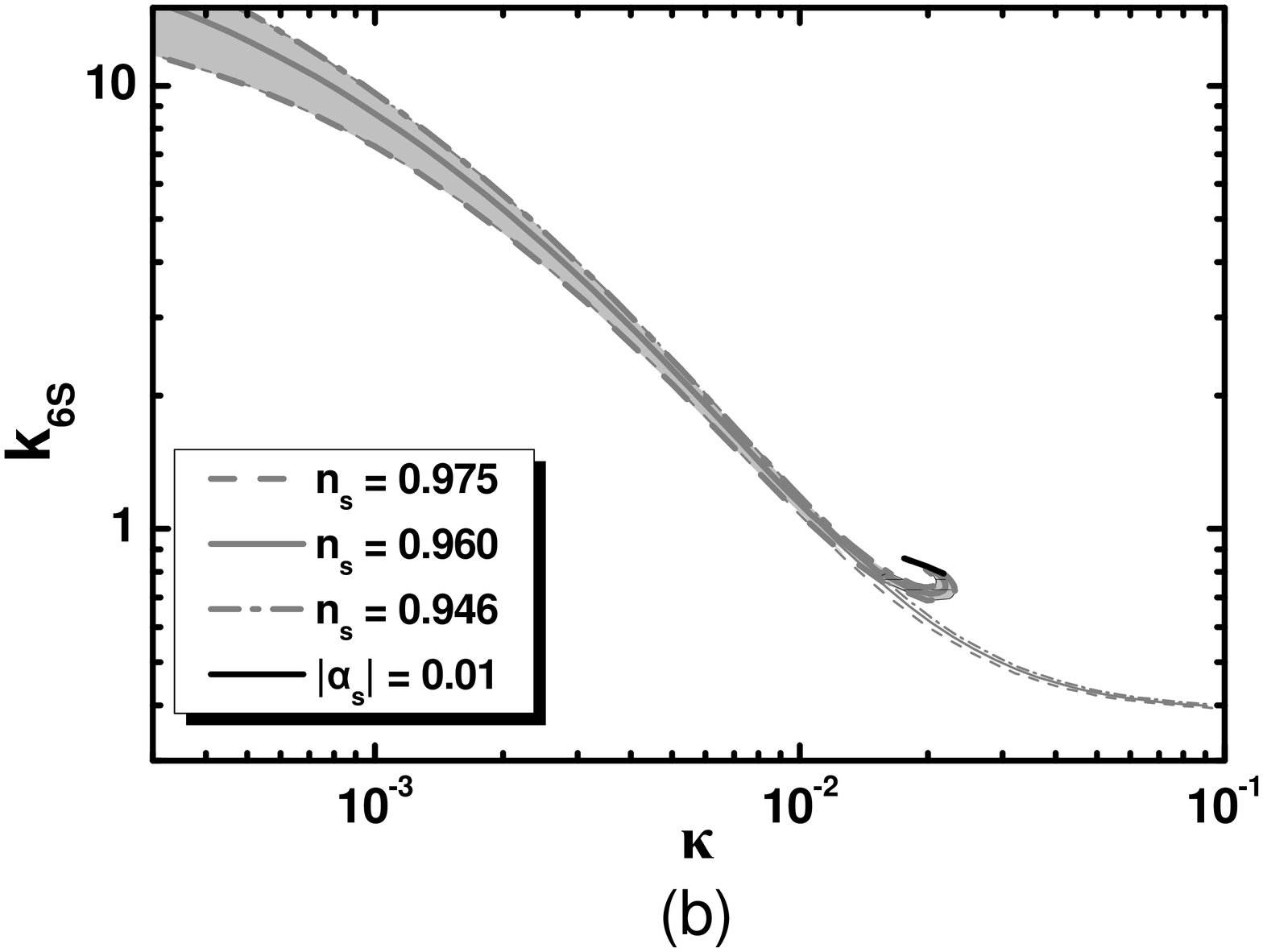}\\
\includegraphics[width=60mm,angle=-90]{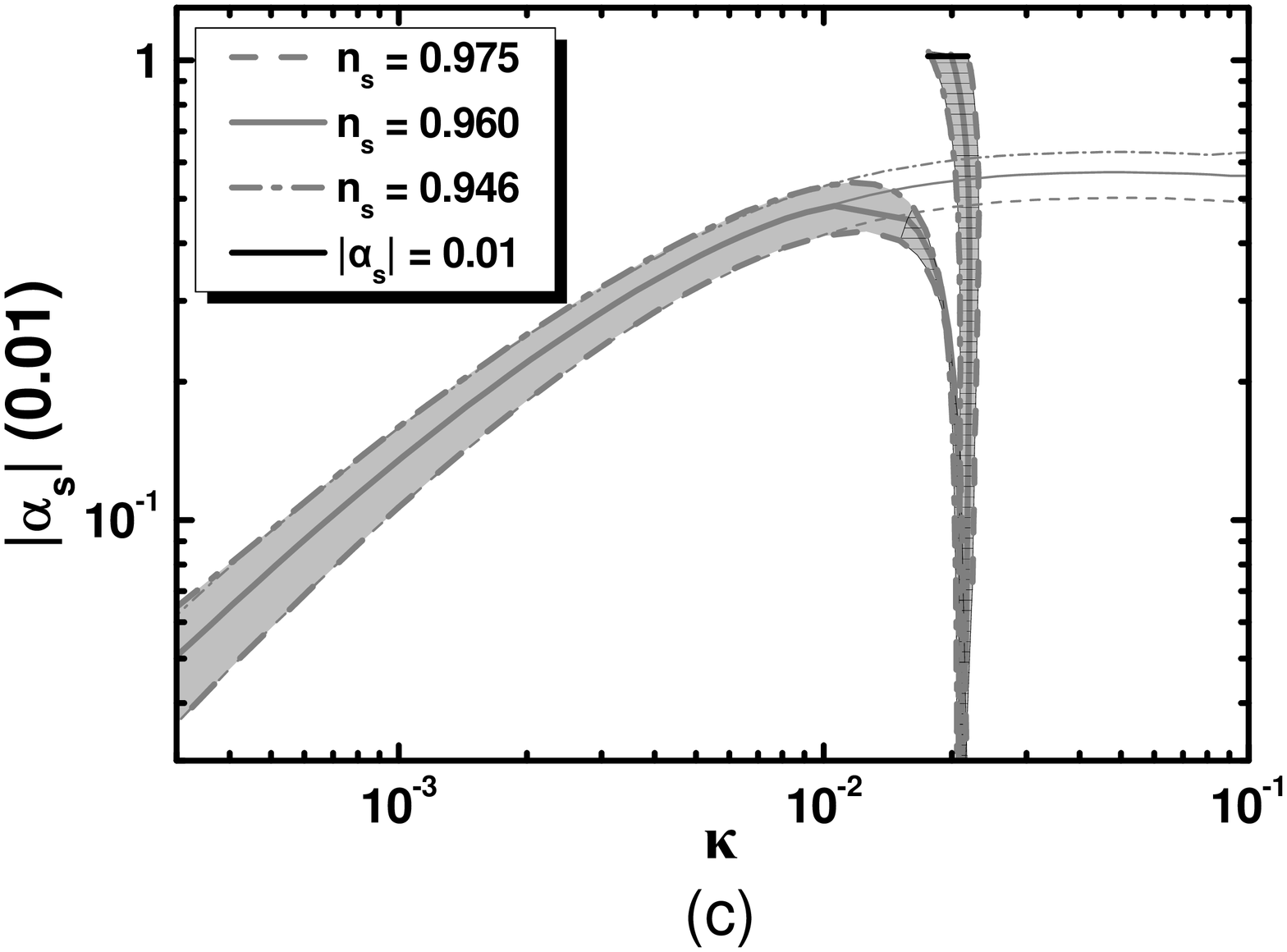}
\includegraphics[width=60mm,angle=-90]{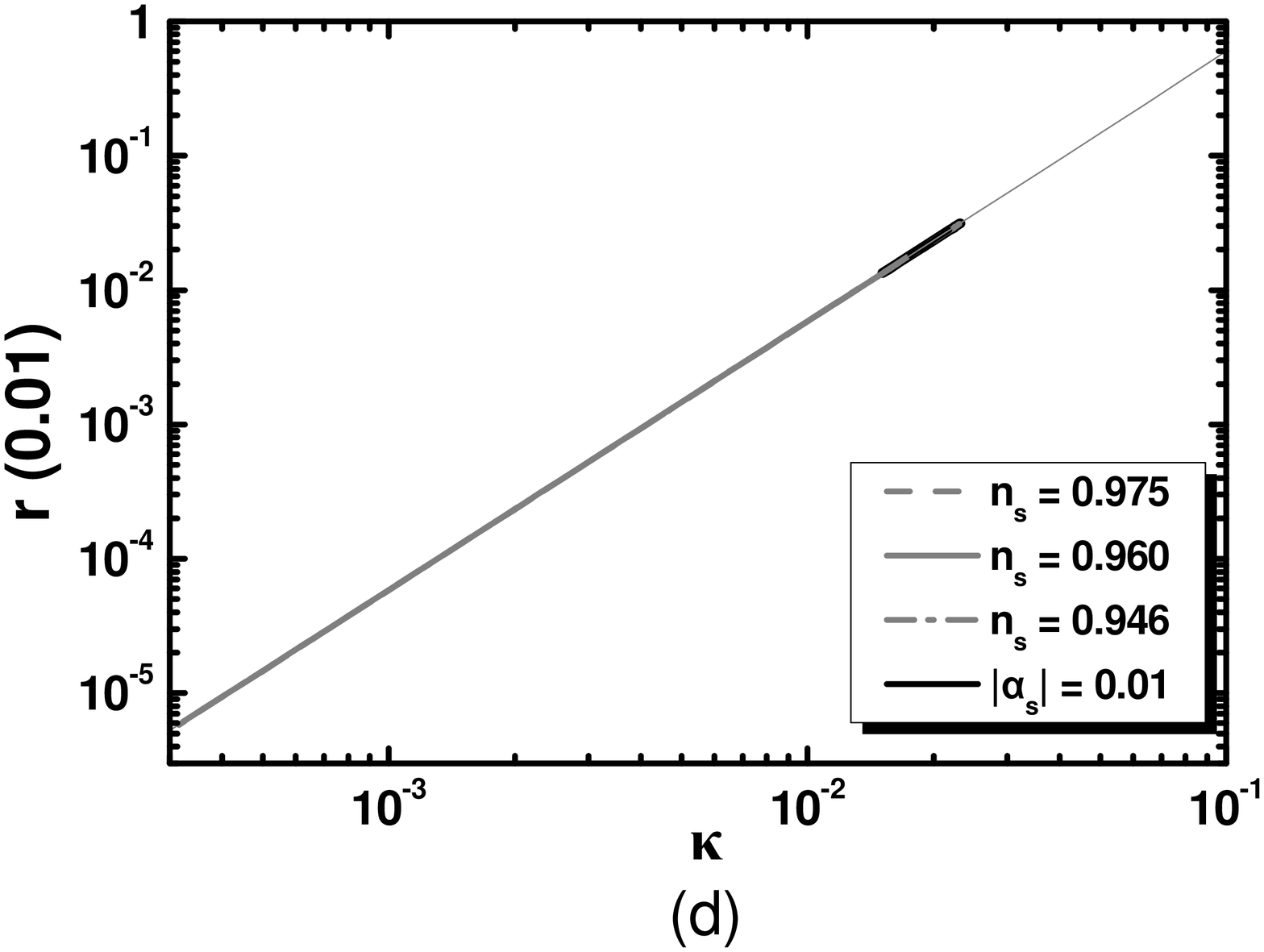}
\caption{\label{fig1}\sl Allowed (lightly gray shaded) region, as
determined by the restrictions 1-6 of \Sref{fhi3}, in the
$\kappa-(-\ks)$ (a), $\kappa-\kss$ (b), $\kappa-|\as|$ (c) and
$\kappa-r$ (d) plane for $\Nr=10$, $\ksss=-1.5,~\kst=-1$ and
$\ksv=0.5$. In the hatched regions $\Vhi$ remains monotonic. The
conventions adopted for the various lines are also shown. The thin
lines are obtained by setting $c_{6K}=c_{8K}=c_{10K}=0$ in
\Eref{Vol}.}
\end{figure*}

The structure of $\Vhi$ is displayed in \Fref{Vhi} where we show
the variation of $\Vhi$ as a function of $\sg$ for $\kp=0.018$ and
$\ks=-0.0443,\kss=0.736,\ksss=-1.5$ (gray line) or
$\ks=-0.0415,\kss=0.656,\ksss=-0.5$ (light gray line). These
parameters yield $\ns=0.96$, $r\simeq0.00019$ and
$\as\simeq0.0054$ $[\as\simeq0.0037]$ (gray [light gray] line).
The values of $\sigma_*/M\simeq19.03$ [$\sigma_*/M\simeq18.4$]
(gray [light gray] line) and $\sigma_{\rm f}/M\simeq1.42$ are also
depicted. In the first case (gray line) $\Vhi$ remains monotonic
due to the larger $|\ksss|$ value employed. Contrarily, $\Vhi$
develops the minimum-maximum structure, in the second case (light
gray line) with the maximum located at $\sg_{\rm
max}/M=26.6\,\{27.2\}$ and the minimum at $\sg_{\rm
min}/M=53.8\,\{63.5\}$ -- the values obtained via
\eqs{sigmamax}{sigmamin} are indicated in curly brackets.  We find
that $\Dex\simeq0.31$.

Confronting FHI with the constraints of Sec.~\ref{fhi3}, we can
identify the allowed regions in the $\kappa-(-\ks)$,
$\kappa-\kss$, $\kappa-|\as|$ and $\kappa-r$ planes -- see
Fig.~\ref{fig1}. The conventions adopted for the various lines are
also shown. In particular, the thick and thin gray dashed
[dot-dashed] lines correspond to $n_{\rm s}=0.975$ [$n_{\rm
s}=0.946$], whereas the thick and thin gray solid lines are
obtained by fixing $n_{\rm s}=0.96$ -- see Eq.~(\ref{nswmap}). The
thick lines are obtained setting $\ksss=-1.5$ which -- together
with the universally selected $\kst$ and $\ksv$ above -- ensures
the fulfilment of restriction 6 of \Sref{fhi3}; the faint lines
correspond to the choice $c_{6K}=c_{8K}=c_{10K}=0$, which does not
ensure the boundedness of $\Vhi$. From the panels (a), (b) and (c)
we see that the thin lines almost coincide with the thick ones for
$\kp\leq0.01$, and then deviate and smoothly approach some
plateau. The regions allowed by imposing the constraints 1-6 of
\Sref{fhi3} are denoted by light gray shading. In the hatched
subregions, requirement 7 is also met. On the other hand, the
regions surrounded by the thin lines are actually the allowed
ones, when only the restrictions 1-5 of \Sref{fhi3} are satisfied.
The various allowed regions are cut at low $\kp$ values since the
required $\kss$ reaches rather high values (of order 10), which
starts looking unnatural. At the other end, \Eref{aswmap} and
$\sigma_*\simeq\mP$ bounds the allowed areas in the case of
bounded or unbounded $\Vhi$ respectively. For both cases, we
remark that $|\ks|$ increases with $\kp$ whereas $\kss$ drops as
$\kappa$ increases. For fixed $\kp$, increasing $|\ks|$ means
decreasing $\kss$. Moreover, $|\ks|$ is restricted to somewhat
small values in order to avoid the well-known
\cite{lectures,review} $\eta$ problem of FHI. On the other hand,
no tuning for $\kss$ is needed since it is of order unity for most
$\kp$ values.

From \sFref{fig1}{c} we observe that for increasing $\kp$ beyond
$0.01$, $|\as|$ corresponding to the bold lines precipitously
drops at $\kp\simeq0.02$, changes sign and rapidly saturates the
bound of \Eref{aswmap} along the thick black solid line. In other
words, for every $\kp$ in the vicinity of $\kp\simeq0.2$ we have
two acceptable $\kss$ values, as shown in \sFref{fig1}{b} with two
different $\as$ values of either sign. Furthermore, from
\sFref{fig1}{d} we remark that $r$ is largely independent of the
$\ns$ value, and so the various types of lines coincide for both
bound and unbounded $\Vhi$. We also see that $r$ increases almost
linearly with $\kp$ and reaches its maximal value which turns out
to be: {\sf\small (i)} $r\simeq 2.9\cdot10^{-5}$ as $\as$
approaches the bound of \Eref{aswmap}, for bounded $V$; {\sf\small
(ii)} $r\simeq 0.01$ as the inequality $\sg_*\leq\mP$ is saturated
for $\ns\simeq0.975$ and unbounded $\Vhi$. Therefore, lifting
restriction 6 of \Sref{fhi3} allows larger $r$. However, non
vanishing $c_{\nu K}$'s perhaps corresponds to a more natural
scenario.

\begin{table}[!b]
\caption{\sl Model parameters and predictions for $\Nr=10$ and
$\ns\simeq0.96$. We take $\kst=-1, \ksv=0.5$ and various
$\ksss$'s.}
\begin{tabular}{c|@{\hspace{0.1cm}}c@{\hspace{0.3cm}}c@{\hspace{0.3cm}}c@{\hspace{0.3cm}}c@
{\hspace{0.3cm}}c@{\hspace{0.3cm}}c@{\hspace{0.3cm}}c} \toprule
{$-\ksss$}&{$\kp$}&{$\sgex/$} &{$\ks$}&{$\kss$}&{$\Dex$}&$\as$&$r$\\
&{$ (10^{-2})$}&$M$&{$(10^{-2})$}&&{$(\%)$}&$(10^{-3})$&$
(10^{-5})$\\\colrule
{$0.5$} &{$0.5$}&{$6.7$}&{$3.46$}&$2.29$&$28$&$3.7$&$1.5$\\
{$0.5$}&{$1$}&{$11.7$}&{$3.94$}&$1.04$&$29$&$5$&$6.6$\\
{$0.5$}&{$2$}&$20.4$&{$4.2$}&{$0.61$}&$-$&$5.3$&$23$\\\colrule
{$1.5$} &{$0.5$}&{$6.7$}&{$3.46$}&$2.29$&$28$&$3.7$&$1.5$\\
{$1.5$}&{$1$}&{$11.5$}&{$3.98$}&$1.1$&$30$&$4.8$&$5.9$\\
{$1.5$}&{$2$}&$23.64$&{$4.68$}&{$0.715$}&$-$&$2.16$&$23.6$\\\colrule
{$2.$} &{$0.5$}&{$6.7$}&{$3.46$}&$2.29$&$28$&$3.7$&$1.5$\\
{$2.$}&{$1$}&{$11.54$}&{$3.98$}&$1.11$&$30$&$4.7$&$6.3$\\
{$2.$}&{$2$}&{$23.4$}&{$5.2$}&$0.785$&$-$&$-8.3$&$23$\\\colrule
\botrule
\end{tabular}
\label{tab1}
\end{table}

We observe that the optimistic restriction 7 in \Sref{fhi3} can be
met in very limited slices of the allowed (lightly gray shaded)
areas, only when the boundedness of $\Vhi$ has been ensured. In
these regions $\sg_*$ also turns out to be rather large ($10M$),
and we therefore observe a mild dependence of our results on
$c_{6K}$ (or $\ksss$). This point is further clarified in
\Tref{tab1} where we list the model parameters and predictions for
$\ns\simeq0.96$, $\Nr=10, \kp=0.005, 0.01, 0.02$ and various
$\ksss$ values. We remark that for $\kp=0.005$ the results are
practically unchanged for varying $\ksss$. The dependence on
$\ksss$ starts to become relevant for $\kp\simeq0.01$ and
crucially affects the results for $\kp=0.02$; here, for $\ksss=-2$
the solution obtained belongs to the branch with $\as<0$ and not
in the branch with $\as>0$, as is the case with $\kp=0.005$ and
$0.01$. Listed is also the quantity $\Dex$ which takes rather
natural values for the selected $\kp$ -- the entries without a
value assigned indicate that $\Vhi$ is a monotonic function of
$\sg$.

\begin{figure}[!t]
\includegraphics[width=60mm,angle=-90]{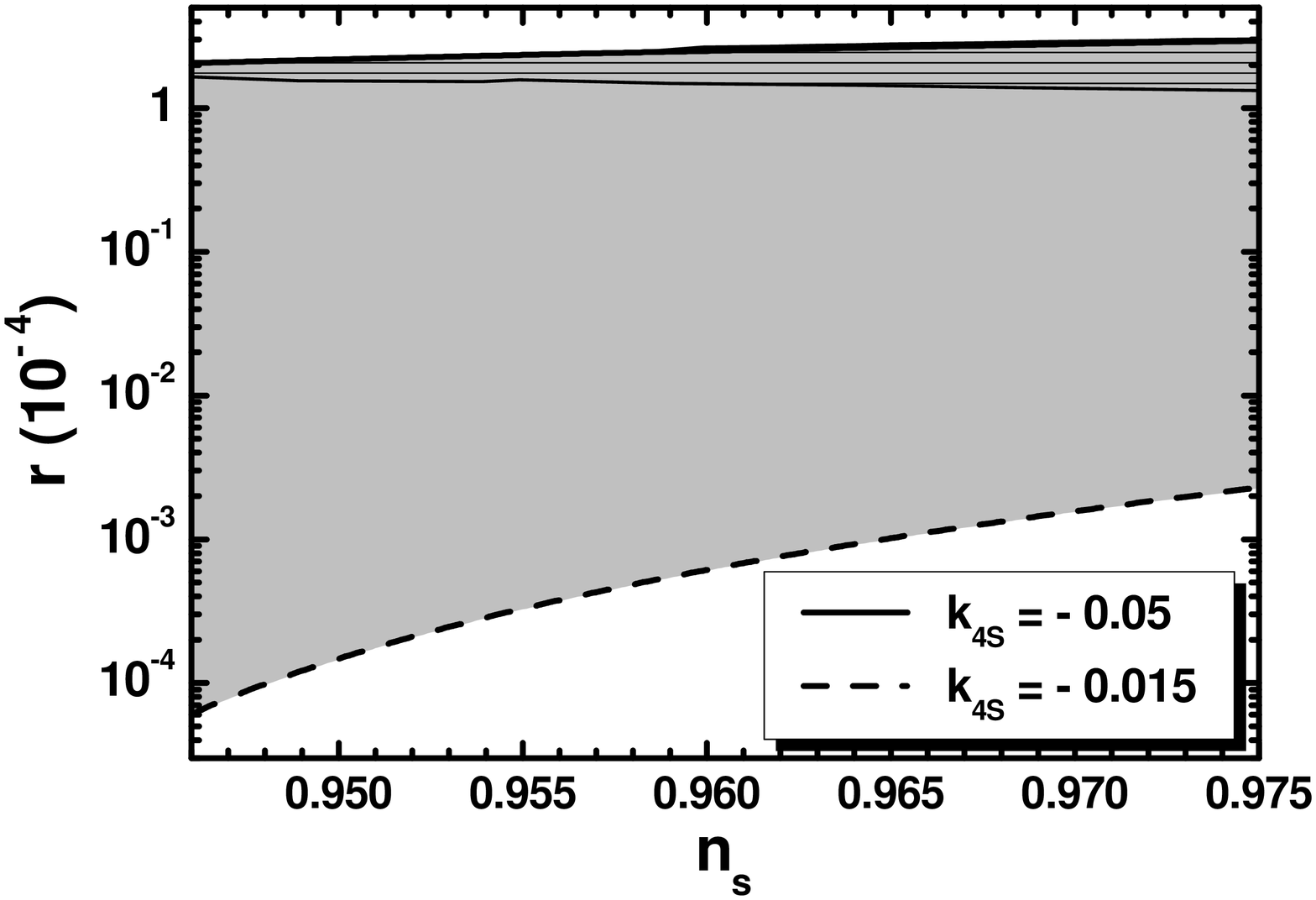}
\includegraphics[width=60mm,angle=-90]{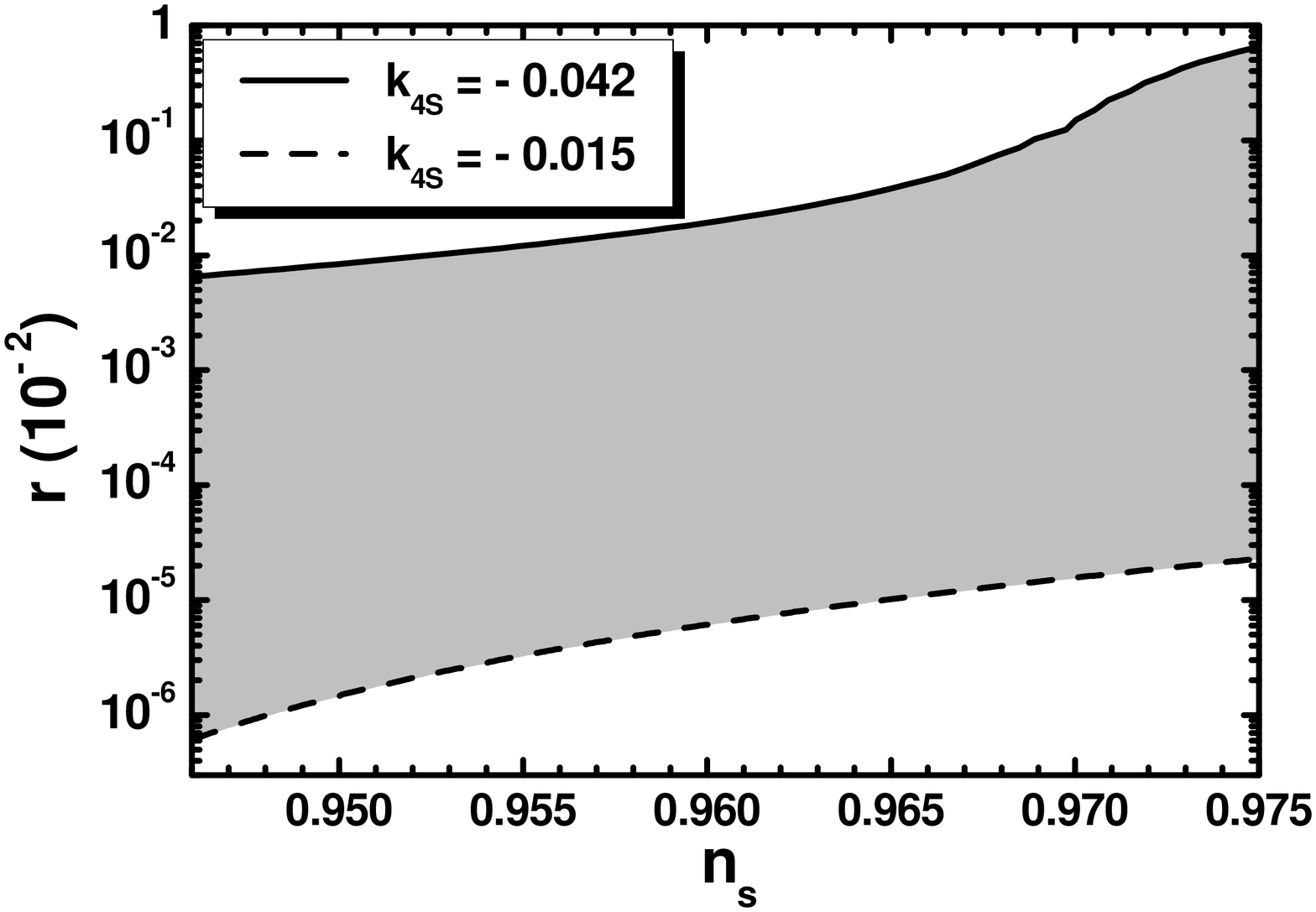}
\vspace*{-0.5cm} \caption{\sl Variation of $r$ as a function of
$\ns$ for $\Nr=10$. We set $\ksss=-1.5$, $\kst=-1$, $\ksv=0.5$,
$\ks=-0.05$ (solid line) and $\ks=-0.015$ (dashed line) for the
upper plot. For the lower plot we set $c_{6K}=c_{8K}=c_{10K}=0$
and $\ks=-0.042$ (solid line) or $\ks=-0.015$ (dashed line). The
shaded region between the two curves is approximately the allowed
region. Hatched is the region in which $\Vhi$ remains
monotonic.}\label{nsr}
\end{figure}

As shown in \sFref{fig1}{a}, $|\ks|$ ranges between about $0.015$
and $0.05$ for the case with bounded $\Vhi$ or $0.042$ for
unbounded $\Vhi$. For each of these $\ks$ values and every $\kp$
in the allowed range found in \Fref{fig1}, we vary $\kss$ in order
to obtain $\ns$ in the observationally favored region of
\Eref{nswmap} and  we extract the resulting $r$. Our results are
presented in \Fref{nsr}, where we display the allowed region in
the $\ns-r$ plane for bounded (upper plot) or unbounded (lower
plot) $\Vhi$. Along the dashed lines of both plots $\kss$ ranges
between $9$ and $26$ whereas along the solid line of the upper
[lower] plot $\kss$ varies between $0.69$ and $0.75$ [$0.39$ and
$1.15$]. From the upper plot we see that the maximal for $r$ is
about $2.9\cdot10^{-4}$ and turns out to be nearly independent of
$\ns$. Interestingly, this value is included in the region with
monotonic $\Vhi$ depicted by the hatched region. From the lower
plot we see that there is a mild dependence of the largest $r$
from $\ns$; thus, the maximal $r=0.006$ is achieved for
$\ns=0.975$. No region with monotonic $\Vhi$ is located in this
case, however.

Summarizing our findings from Figs.~\ref{fig1} and \ref{nsr} for
$\ns$ in the range given by \Eref{nswmap} and imposing the
restrictions 1-7 of \Sref{fhi3}, the various quantities are
bounded as follows:
\beqs\bea \label{rennm2} &&
\{4.9\cdot10^{-2}\}~1.5\lesssim\frac{\kp}{10^{-2}}\lesssim2.3,\\&&
\{1.4\}~4\lesssim\frac{-\ks}{10^{-2}}\lesssim7.95,\\ &&
0.68~\lesssim {\kss}\lesssim0.77~\{10\},\\\label{rennm3}
&&\{5.7\cdot10^{-2}\}~0.4\lesssim
{|\as|\over10^{-2}}\lesssim1,\\\label{rennm4} &&
\{1.7\cdot10^{-3}\}~1.3\lesssim\frac{r}{10^{-4}}\lesssim2.9.
~~~~~~~~~~\eea\eeqs
Note that the limiting values obtained without imposing the
monotonicity of $\Vhi$ -- requirement 7 in \Sref{fhi3} -- are
indicated in curly brackets. In the corresponding region, $\Dex$
ranges between $16$ and $32\%$. As can be deduced from the data of
\Fref{fig1}, $\Dex$ increases with $\kp$'s. Small $\Dex$ values
indicate a second mild tuning (besides the one needed to avoid the
$\eta$ problem), which is however a common feature in the models
of hilltop inflation. The predicted $r$ values are close to the
lowest detectable tensor fraction through cosmic microwave
background polarization \cite{referee}; these are thus virtually
impossible to be observed experimentally. Possibly detectable $r$
values can be achieved if we ignore requirement 6 of \Sref{fhi3}.
Indeed, confining $\ns$ in the range of \Eref{nswmap} we obtain
the following ranges:
\beqs\bea \label{rennm2a} &&
4.9\cdot10^{-3}\lesssim\frac{\kp}{10^{-1}}\lesssim1,\\&&
1.4\lesssim\frac{-\ks}{10^{-2}}\lesssim4.7,\\ && 0.4~\lesssim
{\kss}\lesssim10,\\\label{rennm3a} &&5.7\cdot10^{-1}\lesssim
{|\as|\over10^{-3}}\lesssim6,\\\label{rennm4a} &&
1.4\cdot10^{-5}\lesssim\frac{r}{10^{-2}}\lesssim1.
~~~~~~~~~~\eea\eeqs Obviously, no solutions with monotonic $\Vhi$
are achieved in this case whereas $\Dex$ varies between $16$ and
$29\%$. The maximal $r$ is reached for the maximal $\ns$ in
\Eref{nswmap} and as $\sg_*\sim\mP$.

\begin{table}[!t]
\caption{\sl Model parameters and predictions for $\Nr=2$ and
$\ns\simeq0.96$. We set $\ksss=-1.5,~\kst=-1$ and $\ksv=0.5$.}
\begin{tabular}{c@{\hspace{0.1cm}}c@{\hspace{0.3cm}}c@{\hspace{0.3cm}}c@
{\hspace{0.3cm}}c@{\hspace{0.3cm}}c@{\hspace{0.3cm}}c} \toprule
{$\kp$}&{$\sgex/$} &{$-\ks$}&{$\kss$}&{$\Dex$}&$\as$&$r$\\
{$ (10^{-2})$}&$M$&{$(10^{-2})$}&&{$(\%)$}&$(10^{-4})$&$
(10^{-5})$\\\colrule
{$0.5$} &{$6.4$}&{$3.65$}&{$2.545$}&$28$&$4.1$&$1.5$\\
{$1$}&{$10.4$}&{$4.35$}&{$1.315$}&$30$&$5.7$&$6.2$\\
{$2$}&{$17.3$}&{$5.1$}&{$0.816$}&$35$&$6.3$&$23$\\\botrule
\end{tabular}
\label{tab2}
\end{table}

So far we focused on $\Gfl$, employing $\Nr=10$ in our
investigation. However, our results are not drastically affected
even in the case of $\Glr$ for most values of $\kp$, as can be
inferred by comparing the results (for $\ksss=-1.5$) listed in
Tables~\ref{tab1} and \ref{tab2} where we use $\Nr=10$ and $\Nr=2$
respectively. This signals the fact that the SUGRA corrections to
$\Vhi$ originating from the last term in the sum of \Eref{Vol}
dominate over the radiative corrections which are represented by
$c_{\rm HI}$. The discrepancy between the two results ranges from
$6$ to $20\%$, increasing with $\kp$, and it is essentially
invisible in the plots of \Fref{fig1}. On the other hand, we
observe that in the $\Nr=10$ case the enhanced $c_{\rm HI}$
creates a relatively wider space with monotonic $\Vhi$; this space
is certainly smaller for $\Nr=2$, as shown from our outputs for
$\kp=0.02$.

\section{Conclusions}\label{con}

Inspired by the recently released results by the PLANCK
collaboration on the inflationary observables, we have reviewed
and updated the nonminimal version of SUSY hybrid inflation
arising from F-terms, also referred to as FHI. In our formulation,
FHI is based on a unique renormalizable superpotential, employs an
quasi-canonical \Ka\ and is followed by the spontaneous breaking
at $\Mgut$ of a GUT symmetry which is taken to be $\Glr$ or
$\Gfl$. As suggested first in \cref{rlarge} and further
exemplified in \cref{hinova,alp}, $\ns$ values close to $0.96$ in
conjunction with the fulfilment of \Eref{Mgut} can be accommodated
by considering an expansion of the \Ka\ -- see \Eref{K} -- up to
twelfth order in powers of the various fields with suitable choice
of signs for the coefficients $\ks$ and $\kss$.

Fixing $n_{\rm s}$ at its central value, we obtain
$\{7.8\cdot10^{-2}\}~1.57\lesssim{\kp}/{10^{-2}}\lesssim2.2$ with
$\{2\}~4.2\lesssim{-\ks}/{10^{-2}}\lesssim7.2$ and $0.72~\lesssim
{\kss}\lesssim0.79~\{10\}$, while $|\as|$ and $r$ assume the
values $(\{0.1\}~0.45-1)\cdot10^{-2}$ and
$\lf\{3.5\cdot10^{-3}\}~1.4-1.9\rg\cdot10^{-4}$ respectively --
recall that the limiting values in the curly brackets are achieved
without imposing the monotonicity of $\Vhi$. With a non-monotonic
$\Vhi$, $\Dex$ ranges between $16$ and $30\%$.  It is gratifying
that there is a sizable portion of the allowed parameter space
where $\Vhi$ remains a monotonically increasing function of $\sg$;
thus, unnatural restrictions on the initial conditions for
inflation due to the appearance of a maximum and a minimum of
$\Vhi$ can be avoided. On the other hand, if we do not insist on
the boundedness of $\Vhi$, $\kp$ reaches $0.1$ with $\ks=-0.046$
and $\kss=0.4$ with the resulting $\as$ and $r$ being both
$0.006$, that is close to $0.01$. Finally FHI can be followed by a
successful scenario of non-thermal leptogenesis \cite{lept} for
both $\Ggut$'s considered here -- cf. \cref{flipped,alp}.

\section*{\normalsize\bf\scshape Note Added}\label{note} After the completion of this
work, the {\sc Bicep2} collaboration \cite{gws} recently reported
the discovery of B-mode polarization of the cosmic microwave
background radiation. If this mode is attributed to the primordial
gravity waves predicted by inflation, it implies \cite{gws}
$r=0.16^{+0.06}_{-0.05}$ -- after subtraction of the various dust
models -- which is partially in tension with the WMAP and PLANCK
results \cite{plin,wmap} -- see \Eref{obs4}. Therefore, it is
still premature to exclude any inflationary model with $r$ lower
than the above limit. Moreover, the current data cannot
definitively rule out other sources of gravitational waves -- see
e.g. \cref{dent}. The inflationary models considered in this work
yield $r$ values well below those required by {\sc Bicep2} results
\cite{gws}, especially for inflationary potentials bounded from
below -- see \eqs{rennm4}{rennm4a}. If the {\sc Bicep2} results
are confirmed by other ongoing experiments, the present class of
models defined by the superpotential in \Eref{Whi}, the K\"ahler
potential in \Eref{K} and the theoretical constraint in
\Eref{Mgut} can be categorically excluded.

\acknowledgments  M.C. and Q.S. acknowledge support by the DOE
grant No. DE-FG02-12ER41808. CP acknowledges support from the
Generalitat Valenciana under grant PROMETEOII/2013/017.


\def\ijmp#1#2#3{{\sl Int. Jour. Mod. Phys.}
{\bf #1},~#3~(#2)}
\def\plb#1#2#3{{\sl Phys. Lett. B }{\bf #1}, #3 (#2)}
\def\prl#1#2#3{{\sl Phys. Rev. Lett.}
{\bf #1},~#3~(#2)}
\def\rmp#1#2#3{{Rev. Mod. Phys.}
{\bf #1},~#3~(#2)}
\def\prep#1#2#3{{\sl Phys. Rep. }{\bf #1}, #3 (#2)}
\def\prd#1#2#3{{\sl Phys. Rev. D }{\bf #1}, #3 (#2)}
\def\npb#1#2#3{{\sl Nucl. Phys. }{\bf B#1}, #3 (#2)}
\def\npps#1#2#3{{Nucl. Phys. B (Proc. Sup.)}
{\bf #1},~#3~(#2)}
\def\mpl#1#2#3{{Mod. Phys. Lett.}
{\bf #1},~#3~(#2)}
\def\jetp#1#2#3{{JETP Lett. }{\bf #1}, #3 (#2)}
\def\app#1#2#3{{Acta Phys. Polon.}
{\bf #1},~#3~(#2)}
\def\ptp#1#2#3{{Prog. Theor. Phys.}
{\bf #1},~#3~(#2)}
\def\n#1#2#3{{Nature }{\bf #1},~#3~(#2)}
\def\apj#1#2#3{{Astrophys. J.}
{\bf #1},~#3~(#2)}
\def\mnras#1#2#3{{MNRAS }{\bf #1},~#3~(#2)}
\def\grg#1#2#3{{Gen. Rel. Grav.}
{\bf #1},~#3~(#2)}
\def\s#1#2#3{{Science }{\bf #1},~#3~(#2)}
\def\ibid#1#2#3{{\it ibid. }{\bf #1},~#3~(#2)}
\def\cpc#1#2#3{{Comput. Phys. Commun.}
{\bf #1},~#3~(#2)}
\def\astp#1#2#3{{Astropart. Phys.}
{\bf #1},~#3~(#2)}
\def\epjc#1#2#3{{Eur. Phys. J. C}
{\bf #1},~#3~(#2)}
\def\jhep#1#2#3{{\sl J. High Energy Phys.}
{\bf #1}, #3 (#2)}
\newcommand\jcap[3]{{\sl J.\ Cosmol.\ Astropart.\ Phys.\ }{\bf #1}, #3 (#2)}
\newcommand\njp[3]{{\sl New.\ J.\ Phys.\ }{\bf #1}, #3 (#2)}

\end{document}